\documentclass[11pt,a4paper]{article}
\pdfoutput=1

\usepackage{jheppub}

\usepackage[utf8]{inputenc}
\usepackage[T1]{fontenc}
\usepackage{afterpage}
\usepackage{slashed}

\newcommand{\eVq}{\ensuremath{\text{eV}^2}}
\newcommand{\Dmq}{\Delta m^2}
\newcommand{\Eps}{\varepsilon}
\newcommand{\Epx}{\mathcal{E}}
\newcommand{\diag}{\operatorname{diag}}
\newcommand{\Det}{\operatorname{Det}}
\newcommand{\Tr}{\operatorname{Tr}}

\renewcommand{\Im}{\operatorname{Im}}

\allowdisplaybreaks

\title{On the Determination of Leptonic CP Violation and Neutrino Mass Ordering
  in Presence of Non-Standard Interactions: Present Status}

\author[a]{Ivan Esteban,}
\affiliation[a] {Departament de Fis\'{\i}ca Qu\`antica i
  Astrof\'{\i}sica and Institut de Ciencies del Cosmos, Universitat de
  Barcelona, Diagonal 647, E-08028 Barcelona, Spain}
\emailAdd{ivan.esteban@fqa.ub.edu}

\author[b,a,c]{M.~C.~Gonzalez-Garcia,}
\affiliation[b]{C.N.~Yang Institute for Theoretical Physics, State
  University of New York at Stony Brook, Stony Brook, NY 11794-3840,
  USA}
\affiliation[c]{Instituci\'o Catalana de Recerca i Estudis
  Avan\c{c}ats (ICREA), Pg.\ Lluis Companys 23, 08010 Barcelona,
  Spain}
\emailAdd{maria.gonzalez-garcia@stonybrook.edu}

\author[d]{Michele Maltoni}
\affiliation[d]{Instituto de F\'{\i}sica Te\'orica UAM/CSIC, Calle de
  Nicol\'as Cabrera 13--15, Universidad Aut\'onoma de Madrid,
  Cantoblanco, E-28049 Madrid, Spain}
\emailAdd{michele.maltoni@csic.es}

\abstract{We perform a global analysis of neutrino data in the
  framework of three massive neutrinos with non-standard neutrino
  interactions which affect their evolution in the matter
  background. We focus on the effect of NSI in the present observables
  sensitive to leptonic CP violation and to the mass ordering.  We
  consider complex neutral current neutrino interactions with quarks
  whose lepton-flavor structure is independent of the quark type. We
  quantify the status of the ``hints'' for CP violation, the
  mass-ordering and non-maximality of $\theta_{23}$ in these
  scenarios. We also present a parametrization-invariant formalism for
  leptonic CP violation in presence of a generalized matter potential
  induced by NSI.}


\preprint{IFT-UAM/CSIC-19-65, YITP-SB-2019-10}

\arxivnumber{XXX}

\begin{document}

\maketitle

\section{Introduction}

Experiments measuring the flavor composition of neutrinos produced in
the Sun, in the Earth's atmosphere, in nuclear reactors and in
particle accelerators have established that lepton flavor is not
conserved in neutrino propagation, but it oscillates with a wavelength
which depends on distance and energy. This demonstrates beyond doubt
that neutrinos are massive and that the mass states are non-trivial
admixtures of flavor states~\cite{Pontecorvo:1967fh, Gribov:1968kq},
see Ref.~\cite{GonzalezGarcia:2007ib} for an overview.

The most recent global analysis of the neutrino oscillation data in
Ref.~\cite{Esteban:2018azc} (see also~\cite{deSalas:2017kay,
  Capozzi:2018ubv}) in the context of $3\nu$-mixing provides us with
the determination of the three leptonic mixing angles and the two mass
differences with a $3\sigma/3$ precision of 3\% for $\theta_{12}$, 3\%
for $\theta_{13}$, 9\% for $\theta_{23}$, 5\% for $\Dmq_{21}$ and
2.5\% for $|\Dmq_{31}|$.  Questions still open in the analysis are the
maximality and octant of $\theta_{23}$, the ordering of the mass
eigenstates, and the status of the leptonic CP violating phase
$\delta_\text{CP}$. Some hints in this respect are emerging --~with a
special role played by the long-baseline (LBL) accelerator experiments
T2K~\cite{Abe:2017vif, Abe:2018wpn}, and NOvA~\cite{Adamson:2017gxd,
  NOvA:2018gge}~-- but without a consolidated statistical significance
yet. The latest results show a preference of the normal ordering (NO)
at the 2-3$\sigma$ level, and a best fit for the complex phase at
$\delta_\text{CP} = 215^\circ$ close to maximal CP violation. The
clarification of these unknowns is the main focus of the running LBL
experiments and its precise determination is at the center of the
physics program of the upcoming LBL facilities, in particular the Deep
Underground Neutrino Experiment (DUNE)~\cite{Acciarri:2016ooe} and the
Tokay to HyperKamiokande (T2HK) experiment~\cite{Abe:2015zbg}.

Under the assumption that the Standard Model (SM) is the low energy
effective model of a complete high energy theory, neutrino masses
emerge naturally as the first observable consequence in the form of
the Weinberg operator~\cite{Weinberg:1979sa}, the only dimension five
operator that can be built within the SM particle content.  In this
framework the next operators with observable consequences at low
energies appear at dimension six.  They include four-fermion terms
leading to Non-Standard Interactions (NSI)~\cite{Wolfenstein:1977ue,
  Valle:1987gv, Guzzo:1991hi} between neutrinos and matter (for a
recent review see~\cite{Farzan:2017xzy}).

Neutral Current NSI can modify the forward-coherent scattering
(\textit{i.e.}, at zero momentum transfer) of neutrinos as they
propagate through matter via the so-called Mikheev-Smirnov-Wolfenstein
(MSW) mechanism~\cite{Wolfenstein:1977ue,
  Mikheev:1986gs}. Consequently their effect can be significantly
enhanced in oscillation experiments where neutrinos travel large
regions of matter. Indeed, the global analysis of data from
oscillation experiments in the framework of mass induced oscillations
in presence of NSI currently provides some of the strongest
constraints on the size of the NSI affecting neutrino
propagation~\cite{GonzalezGarcia:2011my, Gonzalez-Garcia:2013usa,
  Esteban:2018ppq}.

In the presence of such NSI, however, the task of exploring leptonic
CP violation in LBL experiments becomes enriched (to the point of
confusion) by the possible coexistence of new sources of CP
violation~\cite{GonzalezGarcia:2001mp}.  Furthermore, the
determination of the mass ordering is jeopardized by the presence of
an intrinsic degeneracy in the relevant oscillation probabilities due
to a new symmetry of the Hamiltonian describing the neutrino evolution
in the modified matter potential~\cite{GonzalezGarcia:2011my,
  Gonzalez-Garcia:2013usa, Bakhti:2014pva, Coloma:2016gei}.
This has resulted in an intense phenomenological activity to quantify
these issues and to devise strategies to clarify them, first in
proposed facilities like the Neutrino
Factory~\cite{Bandyopadhyay:2007kx, Gago:2009ij, Coloma:2011rq} and
most recently in the context of the upcoming
experiments~\cite{Coloma:2015kiu, Masud:2015xva, deGouvea:2015ndi,
  Liao:2016hsa, Huitu:2016bmb, Bakhti:2016prn, Masud:2016bvp,
  C.:2017yqh, Rashed:2016rda, Masud:2016gcl, Blennow:2016etl,
  Ge:2016dlx, Forero:2016ghr, Blennow:2016jkn, Fukasawa:2016lew,
  Liao:2016orc, Deepthi:2016erc, Deepthi:2017gxg, Meloni:2018xnk,
  Flores:2018kwk, Verma:2018gwi, Chatterjee:2018dyd, Masud:2018pig}.

Very interestingly, it has been argued that NSIs can already play a
role in the significance of the ``hints'' mentioned
above~\cite{Forero:2016cmb, Liao:2016bgf}.  In particular in
Ref.~\cite{Forero:2016cmb} it was pointed out the discomforting
possibility of confusing CP conserving NSI with a non-zero value of
$\delta_\text{CP}$ in the analysis of $\nu_e$ and $\bar\nu_e$
appearance results at T2K and NOvA.  Clearly such confusion could lead
to an incorrect claim of the observation of leptonic CP-violation in a
theory which is CP conserving.

We recently performed a global analysis of oscillation data in the
presence of NSI relevant to neutrino propagation in matter in
Ref.~\cite{Esteban:2018ppq}.  For simplicity the analysis in
Ref.~\cite{Esteban:2018ppq} only constrained the CP conserving part of
the Hamiltonian and for consistency the observables most sensitive to
CP violating effects, \textit{i.e.}, $\nu_e$ and $\bar\nu_e$
appearance at LBL experiments, were not included in the fit.
Consequently the issue of the possible confusion between real NSI and
leptonic CP-violation could not be addressed.  Furthermore, under the
simplifying assumptions employed, the results of
Ref.~\cite{Esteban:2018ppq} could not yield any conclusion on the
status of the determination of the ordering of the states in the
presence of NSIs either.

In this paper we extend the analysis in Ref.~\cite{Esteban:2018ppq} to
account for the effect of NSI in the observables sensitive to leptonic
CP violation and to the mass ordering.  Our goal is to quantify the
robustness of the present ``hints'' for these effects in the presence
of NSI \emph{which are consistent with the bounds imposed by the
  CP-conserving observables}. We start by briefly summarizing the
formalism and notation in Sec.~\ref{sec:formalism}. In doing so we
take the opportunity to present a parametrization-invariant formalism
for leptonic CP violation in the presence of a generalized matter
potential induced by NSI.  In Sec.~\ref{sec:analysis} we describe the
strategy employed in the study.  Finally in Sec.~\ref{sec:results} we
present the results, and in Sec.~\ref{sec:summary} we summarize our
conclusions.  We present some detail of the construction of the basis
invariants for CP violation in Appendix~\ref{sec:appendix}.

\section{Formalism}
\label{sec:formalism}

In this work we will consider NSI affecting neutral-current processes
relevant to neutrino propagation in matter. The coefficients
accompanying the relevant operators are usually parametrized in the
form:
\begin{equation}
  \label{eq:NSILagrangian}
  \begin{aligned}
    \mathcal{L}_\text{NSI}
    &= -2\sqrt2 G_F
    \sum_{f,\alpha,\beta} \Eps_{\alpha\beta}^f
    (\bar\nu_\alpha\gamma^\mu P_L\nu_\beta)
    (\bar f\gamma_\mu  f) \,,
    \\
    &= -2\sqrt2 G_F
    \Big[ \sum_{\alpha,\beta} \Eps_{\alpha\beta}
    (\bar\nu_\alpha\gamma^\mu P_L\nu_\beta) \Big] \,
    \Big[ \sum_{f} \xi^f (\bar f\gamma_\mu  f) \Big]
  \end{aligned}
\end{equation}
where $G_F$ is the Fermi constant, $\alpha, \beta$ are flavor indices,
and $f$ is a SM charged fermion. In this notation,
$\Eps_{\alpha\beta}^f$ parametrizes the strength of the vector part of
the new interactions (which are the ones entering the matter
potential) with respect to the Fermi constant, $\Eps_{\alpha\beta}^f
\sim \mathcal{O}(G_X/G_F)$.
In the second line we make explicit that, as in
Ref.~\cite{Esteban:2018ppq}, we assume that the neutrino flavor
structure of the interactions is independent of the charged fermion
type, so that one can factorize $\Eps_{\alpha\beta}^f \equiv
\Eps_{\alpha\beta}\, \xi^f$.

Ordinary matter is composed of electrons ($e$), up quarks ($u$) and
down quarks ($d$). As in Ref.~\cite{Esteban:2018ppq} we restrict
ourselves to non-standard interactions with quarks, so that only
$\xi^u$ and $\xi^d$ are relevant for neutrino propagation.  A global
rescaling of both $\xi^u$ and $\xi^d$ by a common factor can be
reabsorbed into a rescaling of $\Eps_{\alpha\beta}$, therefore only
the \emph{direction} in the $(\xi^u, \xi^d)$ plane is
phenomenologically non-trivial.

In this framework, the evolution of the neutrino and antineutrino
flavor state during propagation is governed by the Hamiltonian:
\begin{equation}
  H^\nu = H_\text{vac} + H_\text{mat}
  \quad\text{and}\quad
  H^{\bar\nu} = ( H_\text{vac} - H_\text{mat} )^* \,,
\end{equation}
where $H_\text{vac}$ is the vacuum part which in the flavor basis
$(\nu_e, \nu_\mu, \nu_\tau)$ reads
\begin{equation}
  \label{eq:Hvac}
  H_\text{vac} = U_\text{vac} D_\text{vac} U_\text{vac}^\dagger
  \quad\text{with}\quad
  D_\text{vac} = \frac{1}{2E_\nu} \diag(0, \Dmq_{21}, \Dmq_{31}) \,.
\end{equation}
Here $U_\text{vac}$ denotes the three-lepton mixing matrix in
vacuum~\cite{Pontecorvo:1967fh, Maki:1962mu, Kobayashi:1973fv} which
we parametrize as~\cite{Coloma:2016gei}, $U_\text{vac} =
R_{23}(\theta_{23}, 0) \, R_{13}(\theta_{13}, 0) \,
R_{12}(\theta_{12}, \delta_\text{CP})$, where $R_{ij}(\theta, \delta)$
is a complex rotation in the $ij$ plane
\begin{equation}
  \label{eq:rij}
  R_{ij}(\theta, \delta) =
  \begin{pmatrix}
    \cos\theta & \sin\theta e^{-i\delta} \\
    -\sin\theta e^{i\delta} & \cos\theta
  \end{pmatrix} \,.
\end{equation}
Explicitly:
\begin{equation}
  \label{eq:Uvac}
  U_\text{vac} =
  \begin{pmatrix}
    c_{12} c_{13}
    & s_{12} c_{13} e^{i\delta_\text{CP}}
    & s_{13}
    \\
    - s_{12} c_{23} e^{-i\delta_\text{CP}} - c_{12} s_{13} s_{23}
    & \hphantom{+} c_{12} c_{23} - s_{12} s_{13} s_{23} e^{i\delta_\text{CP}}
    & c_{13} s_{23}
    \\
    \hphantom{+} s_{12} s_{23} e^{-i\delta_\text{CP}} - c_{12} s_{13} c_{23}
    & - c_{12} s_{23} - s_{12} s_{13} c_{23} e^{i\delta_\text{CP}}
    & c_{13} c_{23}
  \end{pmatrix}
\end{equation}
where $c_{ij} \equiv \cos\theta_{ij}$ and $s_{ij} \equiv
\sin\theta_{ij}$.

If all possible operators in Eq.~\eqref{eq:NSILagrangian} are added to
the SM Lagrangian the matter part $H_\text{mat}$ is a function of the
number densities for the fermions in the matter $N_f(x)$ along the
trajectory:
\begin{equation}
  \label{eq:Hmat}
  H_\text{mat} = \sqrt{2} G_F N_e(x)
  \begin{pmatrix}
    1 + \Epx_{ee}(x) & \Epx_{e\mu}(x) & \Epx_{e\tau}(x) \\
    \Epx_{e\mu}^*(x) & \Epx_{\mu\mu}(x) & \Epx_{\mu\tau}(x) \\
    \Epx_{e\tau}^*(x) & \Epx_{\mu\tau}^*(x) & \Epx_{\tau\tau}(x)
  \end{pmatrix}
\end{equation}
where the ``$+1$'' term in the $ee$ entry accounts for the standard
contribution, and
\begin{equation}
  \label{eq:epx-nsi}
  \Epx_{\alpha\beta}(x) = \sum_{f=u,d}
  \frac{N_f(x)}{N_e(x)} \Eps_{\alpha\beta}^f
  = \big[ \xi^p + Y_n(x) \xi^n \big] \Eps_{\alpha\beta} \,,
\end{equation}
with
\begin{equation}
  \xi^p \equiv 2\xi^u + \xi^d \,,
  \qquad
  \xi^n \equiv \xi^u + 2\xi^d \,,
  \qquad    
  Y_n(x) \equiv \frac{N_n(x)}{N_e(x)}
  = \frac{2N_u(x) - N_d(x)}{N_e(x)}
\end{equation}
describes the non-standard part, which we have written in terms of the
effective couplings of protons ($p$) and neutrons ($n$) after taking
into account that matter neutrality implies $N_e(x) = N_p(x)$.
So the phenomenological framework adopted here is characterized by
nine matter parameters: eight related to the matrix
$\Eps_{\alpha\beta}$ (after neglecting the trace which is irrelevant
for neutrino oscillations) plus the direction in the $(\xi^p, \xi^n)$
plane.

\subsection{NSI-Mass-ordering degeneracy}

When considering the neutrino evolution in the most general matter
potential described above one finds an intrinsic degeneracy as a
consequence of CPT~\cite{GonzalezGarcia:2011my,
  Gonzalez-Garcia:2013usa, Bakhti:2014pva, Coloma:2016gei}.  In brief
CPT symmetry implies that the neutrino evolution is invariant if the
Hamiltonian $H^\nu = H_\text{vac} + H_\text{mat}$ is transformed as
$H^\nu \to -(H^\nu)^*$.  This requires a simultaneous transformation
of both the vacuum and the matter terms. The transformation of
$H_\text{vac}$ implies
\begin{equation}
  \label{eq:osc-deg}
  \begin{aligned}
    \Dmq_{31} &\to -\Dmq_{31} + \Dmq_{21} = -\Dmq_{32} \,,
    \\
    \theta_{12} & \to \pi/2 - \theta_{12} \,,
    \\
    \delta_\text{CP} &\to \pi - \delta_\text{CP}
  \end{aligned}
\end{equation}
which does not spoil the commonly assumed restrictions on the range of
the vacuum parameters ($\Dmq_{21} > 0$ and $0 \leq \theta_{ij} \leq
\pi/2$). Eq.~\eqref{eq:osc-deg} involves a change in the octant of
$\theta_{12}$ as well as a change in the neutrino mass ordering
(\textit{i.e.}, the sign of $\Dmq_{31}$), which is why it has been
called ``generalized mass ordering degeneracy'' in
Ref.~\cite{Coloma:2016gei}. In the SM, this degeneracy is broken
by the SM matter potential and that is what allows for the
determination of the octant of $\theta_{12}$ in solar neutrino
experiments and of the ordering of the states in LBL experiments.  But
once the NSI are included the mass order degeneracy can be recovered
if together with Eq.\eqref{eq:osc-deg} one also exchanges
\begin{equation}
  \label{eq:NSI-deg}
  \begin{aligned}
    \big[ \Epx_{ee}(x) - \Epx_{\mu\mu}(x) \big]
    &\to - \big[ \Epx_{ee}(x) - \Epx_{\mu\mu}(x) \big] - 2  \,,
    \\
    \big[ \Epx_{\tau\tau}(x) - \Epx_{\mu\mu}(x) \big]
    &\to -\big[ \Epx_{\tau\tau}(x) - \Epx_{\mu\mu}(x) \big] \,,
    \\
    \Epx_{\alpha\beta}(x)
    &\to - \Epx_{\alpha\beta}^*(x) \qquad (\alpha \neq \beta) \,,
  \end{aligned}
\end{equation}
see Refs.~\cite{Gonzalez-Garcia:2013usa, Bakhti:2014pva, Coloma:2016gei}.

As seen in Eq.~\eqref{eq:epx-nsi} the matrix $\Epx_{\alpha\beta}(x)$
depends on the chemical composition of the medium, which may vary
along the neutrino trajectory, so that in general the condition in
Eq.~\eqref{eq:NSI-deg} is fulfilled only in an approximate way.  It
becomes exact in several cases, for example if the couplings to
neutrons cancel ($\xi^n = 0$) so that the position-dependent term in
Eq.~\eqref{eq:epx-nsi} disappears.
More interestingly the mass ordering degeneracy is also recovered when
the neutron/proton ratio, $Y_n(x)$, is constant along the entire
neutrino propagation path. This is the case for long-baseline
experiments, either in accelerators or reactors, as in the Earth the
neutron/proton ratio $Y_n(x)$ which characterize the matter chemical
composition can be taken to be constant to very good approximation.
It is also a reasonably good approximation for atmospheric neutrino
experiments.  The PREM model~\cite{Dziewonski:1981xy} fixes $Y_n =
1.012$ in the mantle and $Y_n = 1.137$ in the core, so that for
atmospheric and LBL neutrino experiments one can set it to an average
value $Y_n^\oplus= 1.051$ all over the Earth.  Hence for those
experiments one gets $\Epx_{\alpha\beta}(x) \equiv
\Eps_{\alpha\beta}^\oplus$ with:
\begin{equation}
  \label{eq:eps-earth}
  \Eps_{\alpha\beta}^\oplus
  = \big( \xi^p + Y_n^\oplus \xi^n \big)\, \Eps_{\alpha\beta} \,.
\end{equation}
In other words, within this approximation the analysis of atmospheric
and LBL neutrinos holds for any combination of NSI with up and down
quarks (and also with electrons if it had been considered) and it can
be performed in terms of the effective NSI couplings
$\Eps_{\alpha\beta}^\oplus$, which play the role of phenomenological
parameters. In particular, the best-fit value and allowed ranges of
$\Eps_{\alpha\beta}^\oplus$ are independent of the $(\xi^p, \xi^n)$
couplings.

\subsection{New sources of CP violation}

Another obvious consequence of the effect of NSI in the matter
potential is the appearance of new sources of CP violation. As seen in
Eqs.~\eqref{eq:Uvac} and~\eqref{eq:Hmat} the Hamiltonian contains now
four possible physical phases which in that parametrization are the
Dirac CP-phase in $U_\text{vac}$ and the three arguments of the
off-diagonal elements in $H_\text{mat}$.  We emphasize that only one
of the phases can be assigned to the vacuum part of the Hamiltonian
and only one can be assigned to the matter Hamiltonian in a basis
independent form. The other two phases are relative phases between
both pieces of the Hamiltonian.  This is more transparent if one uses
the parametrization of Ref.~\cite{GonzalezGarcia:2011my} which for the
relevant part of the NSI matrix in Eq.~\eqref{eq:NSILagrangian} reads:
\begin{equation}
  \label{eq:EpsMic}
  \Eps =
  \begin{pmatrix}
    \Eps_{ee} \hphantom{e^{-i\phi_{xx}}}
    & |\Eps_{e\mu}| e^{i\phi_{e\mu}\hphantom{+}}
    & |\Eps_{e\tau}| e^{i\phi_{e\tau}}
    \\
    |\Eps_{e\mu}| e^{-i\phi_{e\mu}}
    & \Eps_{\mu\mu} \hphantom{e^{-i\phi_{xx}}}
    & |\Eps_{\mu\tau}| e^{i\phi_{\mu\tau}}
    \\
    |\Eps_{e\tau}| e^{-i\phi_{e\tau}}
    & |\Eps_{\mu\tau}| e^{-i\phi_{\mu\tau}}
    & \Eps_{\tau\tau} \hphantom{e^{i\phi_{xx}}}
  \end{pmatrix}
  \equiv
  Q_\text{rel} U_\text{mat} D_\Eps
  U_\text{mat}^\dagger Q_\text{rel}^\dagger
  - A
  \begin{pmatrix}
    1 & 0 & 0 \\
    0 & 0 & 0 \\
    0 & 0 & 0
  \end{pmatrix}
\end{equation}
with
\begin{equation}
  \begin{aligned}
    Q_\text{rel} &= \diag\big(
    e^{i\alpha_1}, e^{i\alpha_2}, e^{-i\alpha_1 -i\alpha_2} \big),
    \\
    U_\text{mat} &= R_{12}(\varphi_{12},0) R_{13}(\varphi_{13},0)
    {R}_{23}(\varphi_{23}, \delta_\text{NS}) \,,
    \\
    D_\Eps &= \diag(\Eps_1, \Eps_2, \Eps_3)
  \end{aligned}
\end{equation}
and $A$ is a real number. The advantage of this formalism is that,
when propagating in a medium with constant chemical composition (for
example, the Earth with $Y_n(x) = Y_n^\oplus$), it is possible to
choose $A$ in such a way that the matrix $H_\text{mat}$ mimics the
structure of the vacuum term in Eq.~\eqref{eq:Hvac}:
\begin{equation}
  \label{eq:HmatGen}
  H_\text{mat} = Q_\text{rel} U_\text{mat} D_\text{mat}
  U_\text{mat}^\dagger Q_\text{rel}^\dagger
  \quad\text{with}\quad
  D_\text{mat} \propto N_e(x)\, D_\Eps \,.
\end{equation}
Written in this form, it is explicit that the two phases $\alpha_1$
and $\alpha_2$ included in $Q_\text{rel}$ are not a feature of
neutrino-matter interactions, but rather a relative feature between
the vacuum and matter terms.

To further illustrate this point we can introduce four basis
invariants to characterize leptonic CP violation in the presence of
non-standard neutrino interactions as described in the
appendix~\ref{sec:appendix}.  They are built as products of hermitian
matrices, concretely the charged lepton mass-squared matrix $S^\ell =
M^\ell {M^\ell}^\dagger$, the neutrino mass-squared matrix $S^\nu =
M^\nu {M^\nu}^\dagger$, and the $\Eps$ matrix.

One of them can be chosen as to contain only parameters which appear
in the neutrino evolution in vacuum, so it gives a measure of leptonic
CP violation in neutrino oscillations in vacuum:
\begin{multline}
  \label{eq:Jvac}
  \Im\Tr\Big( (S^\ell)^2\, (S^\nu)^2\, S^\ell\, S^\nu \Big)
  = \frac{2}{i} \Det [S^\ell, S^\nu]
  = \frac{1}{4} v(m_e,m_\mu,m_\tau)
  \Dmq_{21}\, \Dmq_{31}\, \Dmq_{23}
  \\
  \times \sin(2\theta_{23}) \sin(2\theta_{12})
  \sin(\theta_{13}) \cos^2(\theta_{13})
  \sin(\delta_\text{12} - \delta_\text{13} + \delta_\text{23}) \,,
\end{multline}
where the third equality shows its form in the charged lepton mass
basis, with $v(m_e, m_\mu, m_\tau) = (m_\tau^2-m_\mu^2)
(m_\tau^2-m_e^2) (m_\mu^2-m_e^2)$.  In writing this expression we have
kept $U_\text{vac} = R_{23}(\theta_{23}, \delta_{23})\cdot
R_{13}(\theta_{13}, \delta_{13})\cdot R_{12}(\theta_{12},
\delta_{12})$ to explicitly show the physical phase appearing in this
invariant $\delta_\text{CP}\equiv\delta_\text{12} - \delta_\text{13} +
\delta_\text{23}$ and which, for convenience, in our parametrization
in Eq.~\eqref{eq:Uvac} we have kept as attached to the 12 rotation by
setting $\delta_{13} = \delta_{23} = 0$.  Equation~\eqref{eq:Jvac}
corresponds to the usual Jarlskog invariant for the leptonic sector.

A second invariant can be chosen to be the one characterizing CP
violation in neutrino propagation in matter in the $E_\nu\rightarrow
\infty $ limit:
\begin{multline}
  \label{eq:Jmat}
  \Im\Tr\Big( (S^\ell)^2\, (\Eps)^2\, S^\ell\, \Eps \Big) =
  \frac{2}{i} \Det [S^\ell, \Eps]= v(m_e,m_\mu,m_\tau)\,
  \Im(\Eps_{e\mu}\, \Eps_{\mu\tau}\, \Eps_{\tau e})
  \\
  = v(m_e,m_\mu,m_\tau)\,
  |\Eps_{e\mu}|\, |\Eps_{e\tau}|\, |\Eps_{\mu\tau}|\,
  \sin(\phi_{e\mu} - \phi_{e\tau} + \phi_{\mu\tau})
  \\
  = \frac{1}{4} v(m_e,m_\mu,m_\tau)
  \Delta\Eps_{13}\, \Delta\Eps_{23}\, \Delta\Eps_{12}
  \sin(2\varphi_{23}) \sin(2\varphi_{12})
  \sin(\varphi_{13}) \cos^2(\varphi_{13})
  \sin(\delta_\text{NS})
\end{multline}
with $\Delta\Eps_{ij} \equiv \Eps_i - \Eps_j$. The last two lines
give its explicit expression in the charged lepton mass basis, and in
particular in the last line we have used a parametrization for
$\Eps_{\alpha\beta}$ as in Eq.~\eqref{eq:EpsMic}.

The other two basis invariants involve both $\Eps$ and $S^\nu$ and can
be formed by two combinations of the rephasing invariants $\Im\big(
\Eps_{\alpha\beta} \,S^\nu_{\beta\alpha} \big)$ for
${\alpha\beta}={e\mu},\,{e\tau},\,{\mu\tau}$ as shown, for example,
in Eqs.\eqref{eq:appinv3} and~\eqref{eq:appinv4}.  In the charged
lepton mass basis they read:
\begin{align}
  \label{eq:inv3}
  \begin{split}
    \Im\big(\Eps_{e\mu} S^\nu_{\mu e} \big)
    &= \frac{1}{2} \cos \theta_{13} \Eps_{e\mu} \big[
      \Dmq_{21} \cos\theta_{23} \sin 2\theta_{12}
      \sin(\delta_{12} + \phi_{e\mu})
      \\
      &\quad + (2 \Dmq_{31} - \Dmq_{21} + \Dmq_{21} \cos 2 \theta_{12})
      \sin\theta_{13} \sin\theta_{23}
      \sin(\delta_{13} - \delta_{23} + \phi_{e\mu}) \big]
    \\
    &= -\frac{1}{4} \cos \theta_{13} \cos \varphi_{13}
    \\
    &\quad \times \big[ \Delta\Eps_{12} \cos\varphi_{23} \sin 2\varphi_{12}
      + \big( \Delta\Eps_{13} + \Delta\Eps_{23}
      + \Delta\Eps_{12} \cos 2\varphi_{12} \big)
      \sin\varphi_{13} \sin\varphi_{23} \big]
    \\
    &\quad \times \big[ \Dmq_{21} \cos\theta_{23} \sin 2\theta_{12}
      \sin(\alpha_1 - \alpha_2 + \delta_{12})
      \\
      &\quad + (\Dmq_{31} + \Dmq_{32} + \Dmq_{21} \cos 2\theta_{12})
      \sin\theta_{13} \sin\theta_{23}
      \sin(\alpha_1 - \alpha_2 + \delta_{13} - \delta_{23}) \big]
    \\
    &\quad + \mathcal{O}(\delta_\text{NS}) \,,
  \end{split}
  \\
  \label{eq:inv4}
  \begin{split}
    \Im\big(\Eps_{e\tau} S^\nu_{\tau e} \big)
    &= \frac{1}{2} \cos\theta_{13} \Eps_{e\tau} \big[
      -\Dmq_{21} \sin\theta_{23} \sin 2\theta_{12}
      \sin(\delta_{12} + \delta_{23} + \phi_{e\tau})
      \\
      &\quad + (\Dmq_{31} + \Dmq_{32} + \Dmq_{21}\cos 2\theta_{12})
      \sin\theta_{13} \cos\theta_{23} \sin(\delta_{13} + \phi_{e\tau}) \big]
    \\
    &= \frac{1}{4} \cos\theta_{13} \cos\varphi_{13}
    \\
    &\quad \times \big[
      - \Delta\Eps_{12} \sin\varphi_{23} \sin 2\varphi_{12}
      + \big(\Delta\Eps_{13} + \Delta\Eps_{23}
      + \Delta\Eps_{12} \cos 2\varphi_{12} \big)
      \sin\varphi_{13} \cos\varphi_{23} \big]
    \\
    &\quad \times \big[
      \Dmq_{21} \sin\theta_{23} \sin 2\theta_{12}
      \sin(2\alpha_1 + \alpha_2 + \delta_{12} + \delta_{23})
      \\
      &\quad -(\Dmq_{31} + \Dmq_{32} + \Dmq_{21}\cos 2\theta_{12})
      \sin\theta_{13} \cos\theta_{23}
      \sin(2\alpha_1 + \alpha_2 + \delta_{13}) \big]
    \\
    &\quad + \mathcal{O}(\delta_\text{NS}) \,.
  \end{split}
\end{align}
and $\Im\big(\Eps_{\mu\tau} S^\nu_{\tau\mu} \big)$ can be written in
terms of the two above using the equality in
Eq.~\eqref{eq:nonIndependentCondition}. In the last expression of the
two equations above we show the explicit form of these invariants in
the parametrization in Eq.~\eqref{eq:EpsMic} only for
$\delta_\text{NS} = 0$ for simplicity.

Unlike for the case of the invariants in Eqs.~\eqref{eq:Jvac}
and~\eqref{eq:Jmat}, there is not a clear physical set up which could
single out the contribution from Eq.~\eqref{eq:inv3} and
Eq.~\eqref{eq:inv4} (or any combination of those) to a leptonic CP
violating observable.  From their explicit expressions one also finds
that setting the $\alpha_i$ phases in $Q_\text{rel}$ as well as
$\delta_\text{NS}$ to zero is not enough to cancel those
``matter-vacuum interference'' CP invariants. On the contrary, setting
those phases to zero one reintroduces a dependence on the phase
convention for the phases in the vacuum mixing matrix.

Admittedly the discussion above is only academic for the
quantification of the effects induced by the NSI matter potential on
neutrino propagation, because the relevant probabilities cannot be
expressed in any practical form in terms of these basis invariants and
one is forced to work in some specific parametrization. What these
basis invariants clearly illustrate is that in order to study the
possible effects (in experiments performed in matter) of NSI on the
determination of the phase which parametrizes CP violation in vacuum
\emph{without introducing an artificial basis dependence}, one needs
to include in the analysis the most general complex NSI matter
potential containing \emph{all} the three additional arbitrary phases.

\section{Analysis Framework}
\label{sec:analysis}

As mentioned in the introduction this work builds upon the results of
our recent comprehensive global fit performed in the framework of
three-flavor oscillations plus NSI with quarks~\cite{Esteban:2018ppq}
to which we refer for the detailed description of methodology and data
included.  In principle, for each choice of the $(\xi^p, \xi^n)$
couplings the analysis depends on six oscillations parameters plus
eight NSI parameters, of which five are real and three are phases. To
keep the fit manageable in Ref.~\cite{Esteban:2018ppq} only real NSI
were considered and $\Dmq_{21}$ effects were neglected in the fit of
atmospheric and long-baseline experiments. This rendered such analysis
independent of the CP phase in the leptonic mixing matrix and of the
ordering of the states.

In this work we extend our previous analysis to include the effect of
the four CP-phases in the Hamiltonian as well as the $\Dmq_{21}$
effects, in particular where they are most relevant which is the fit
of the LBL experiments.  In order to do so while still maintaining the
running time under control, we split the global $\chi^2$ in a part
containing the data from the long-baseline experiments MINOS, T2K and
NOvA (accounting for both appearance and disappearance data in
neutrinos and antineutrino modes), for which both the extra phases and
the interference between $\Dmq_{21}$ and $\Dmq_{31}$ are properly
included, and a part containing CP-conserving observables where the
complex phases can be safely neglected and are therefore implemented
as described in Ref.~\cite{Esteban:2018ppq}. In what follows we label
as ``OTH'' (short for ``others'') these non-LBL observables which
include the results from solar neutrino
experiments~\cite{Cleveland:1998nv, Kaether:2010ag,
  Abdurashitov:2009tn, Hosaka:2005um, Cravens:2008aa, Abe:2010hy,
  sksol:nakano2016, Aharmim:2011vm, Bellini:2011rx, Bellini:2008mr,
  Bellini:2014uqa}, from the KamLAND reactor
experiment~\cite{Gando:2013nba}, from medium-baseline (MBL) reactor
experiments~\cite{dc:cabrera2016, An:2016ses, reno:eps2017}, from
atmospheric neutrinos collected by IceCube~\cite{TheIceCube:2016oqi}
and its sub-detector DeepCore~\cite{Aartsen:2014yll}, and from our
analysis of Super-Kamiokande (SK) atmospheric
data~\cite{Wendell:2014dka}.\footnote{As in
  Ref.~\cite{Esteban:2018ppq} we include here our analysis of SK
  atmospheric data in the form of the ``classical'' samples of
  $e$-like and $\mu$-like events (70 energy and zenith angle bins). As
  discussed in Ref.~\cite{Esteban:2018azc} with such analysis in the
  framework of standard 3-nu oscillations we cannot reproduce the
  sensitivity to subdominant effects associated with the mass ordering
  and $\delta_\text{CP}$ found by SK in their analysis of more
  dedicated samples~\cite{Abe:2017aap}. For that reason we include SK
  atmospheric data but only as part of the ``OTH'' group.}  Also, in
order to quantify the impact on the fit of the data on coherent
neutrino--nucleus scattering collected by the COHERENT
experiment~\cite{Akimov:2017ade} we will show the results after
including COHERENT as part of OTH as well.  It should be noted,
however, that COHERENT bounds are only applicable for models where the
mediator responsible for the NSI is heavier than about 10~MeV, as
discussed in Ref.~\cite{Esteban:2018ppq}.

Schematically, if we denote by $\vec{w}$ the five real oscillation
parameters (\textit{i.e.}, the two mass differences and the three
mixing angles), by $\eta$ the direction in the $(\xi^p, \xi^n)$ plane,
by $|\Eps_{\alpha\beta}|$ the five real components of the neutrino
part of the NSI parameters (two differences of the three diagonal
entries of $\Eps_{\alpha\beta}$, as well as the modulus of the three
non-diagonal entries\footnote{More precisely, in our analysis of OTH
  experiments we consider both the modulus and the sign of the
  non-diagonal $\Eps_{\alpha\beta}$, \textit{i.e.}, we explicitly
  account for the all the CP-conserving values of the three phases:
  $\phi_{\alpha\beta} = 0,\pi$. However, in the construction of
  $\chi^2_\text{OTH}$ these signs are marginalized, so that only the
  modulus $|\Eps_{\alpha\beta}|$ is correlated with
  $\chi^2_\text{LBL}$.}), and by $\phi_{\alpha\beta}$ the three phases
of the non-diagonal entries of $\Eps_{\alpha\beta}$, we split the
global $\chi^2$ for the analysis as
\begin{equation}
  \chi^2_\text{GLOB}(\vec{w}, \delta_\text{CP},
  |\Eps_{\alpha\beta}|, \phi_{\alpha\beta}, \eta)
  = \chi^2_\text{OTH}(\vec{w}, |\Eps_{\alpha\beta}|, \eta)
  + \chi^2_\text{LBL}(\vec{w}, \delta_\text{CP},
  |\Eps_{\alpha\beta}|, \phi_{\alpha\beta}, \eta)
\end{equation}
so $\chi^2_\text{OTH}$ and $\chi^2_\text{LBL}$ depend on $5+5+1=11$
and $5+1+5+3+1 = 15$ parameters, respectively.

To make the analysis in such large parameter space treatable, we
introduce a series of simplifications. First, we notice that in MBL
reactor experiments the baseline is short enough to safely neglect the
effects of the matter potential, so that we have:
\begin{equation}
  \chi^2_\text{OTH}(\vec{w}, |\Eps_{\alpha\beta}|, \eta)
  = \chi^2_\text{SOLAR+KAMLAND+ATM}(\vec{w}, |\Eps_{\alpha\beta}|, \eta)
  + \chi^2_\text{MBL-REA}(\vec{w}) \,.
\end{equation}
Next, we notice that in LBL experiments the sensitivity to
$\theta_{12}$, $\Dmq_{21}$ and $\theta_{13}$ is marginal compared to
solar and reactor experiments; hence, in $\chi^2_\text{LBL}$ we can
safely fix $\theta_{12}$, $\theta_{13}$ and $\Dmq_{21}$ to their best
fit value as determined by the experiments included in
$\chi^2_\text{OTH}$. However, in doing so we must notice that, within
the approximations used in the construction of $\chi^2_\text{OTH}
(\vec{w}, |\Eps_{\alpha\beta}|, \eta)$, there still remains the effect
associated to the NSI/mass-ordering degeneracy which leads to the
appearance of a new solution in the solar sector with a mixing angle
$\theta_{12}$ in the second octant, the so-called LMA-Dark
(LMA-D)~\cite{Miranda:2004nb} solution. Although LMA-D is not totally
degenerate with LMA, due to the variation of the matter chemical
composition along the path travelled by solar neutrinos, it still
provides a good fit to the data for a wide range of quark couplings,
as found in Ref.~\cite{Esteban:2018ppq}.  Concretely, after
marginalization over $\eta$ we get that the parameter region
containing the LMA-D solution lies at
\begin{equation}
  \label{eq:dclmad}
  \chi^2_\text{OTH,LMA-D} - \chi^2_\text{OTH,LMA} = 3.15 \,.
\end{equation}
Therefore, when marginalizing over $\theta_{12}$ we consider two
distinct parts of the parameter space, labelled by the tag ``REG'':
one with $\theta_{12}<45^\circ$, which we denote as $\text{REG} =
\text{LIGHT}$, and one with $\theta_{12}>45^\circ$, which we denote by
$\text{REG} = \text{DARK}$.  Correspondingly, the fixed value of
$\theta_{12}$ used in the construction of $\chi^2_\text{LBL,REG}$ is
the best fit value within either the LMA or the LMA-D region:
$\sin^2\bar\theta_{12}^\textsc{light} = 0.31$ or
$\sin^2\bar\theta_{12}^\textsc{dark} = 0.69$, respectively. The best
fit values for the other two oscillation parameters fixed in
$\chi^2_\text{LBL,REG}$ are the same for LMA and LMA-D:
$\Delta\bar{m}^2_{21} = 7.4\times 10^{-5}~\eVq$ and
$\sin^2\bar\theta_{13} = 0.0225$.

Further simplification arises from the fact that for LBL experiments
the dependence on the NSI neutrino and quark couplings enters only via
the effective Earth-matter NSI combinations
$\Eps_{\alpha\beta}^\oplus$ defined in Eq.~\eqref{eq:eps-earth}.  It
is therefore convenient to project also $\chi^2_\text{OTH}$ over these
combinations, and to marginalize it with respect to $\eta$. In
addition we neglect the small correlations introduced by the common
dependence of the atmospheric experiments in $\chi^2_\text{OTH}$ and
the LBL experiments on $\Dmq_{31}$ and $\theta_{23}$, and we also
marginalize the atmospheric part of $\chi^2_\text{OTH}$ over these two
parameters.  This means that in our results we do not account for the
information on $\Dmq_{31}$ and $\theta_{23}$ arising from atmospheric
experiments, however we keep the information on $\Dmq_{31}$ from MBL
reactor experiments. With all this, we can define a function
$\chi^2_\text{OTH,REG}$ depending on six parameters:
\begin{equation}
  \chi^2_\text{OTH,REG} \big( \Dmq_{31}, |\Eps^\oplus_{\alpha\beta}| \big)
  \equiv
  \min_{\substack{\eta,\, \theta_{12} \in \,\textsc{reg}\\
      \theta_{13}, \theta_{23}, \Dmq_{21}}}
  \chi^2_\text{OTH}\big( \vec{w},\,
  |\Eps_{\alpha\beta}^\oplus| \big/ [\xi^p + Y_n^\oplus \xi^n],\, \eta
  \big)
\end{equation}
while our final global $\chi^2_\text{GLOB,REG}$ is a function of
eleven parameters which takes the form
\begin{multline}
\label{eq:chi2global}
  \chi^2_\text{GLOB,REG} \big( \theta_{23}, \Dmq_{31},
  \delta_\text{CP}, |\Eps_{\alpha\beta}^\oplus|, \phi_{\alpha\beta} \big)
  = \chi^2_\text{OTH,REG} \big( \Dmq_{31}, |\Eps^\oplus_{\alpha\beta}| \big)
  \\
  + \chi^2_\text{LBL,REG} \big( \theta_{23}, \Dmq_{31}, \delta_\text{CP},
  |\Eps_{\alpha\beta}^\oplus|, \phi_{\alpha\beta}
  \,\big\Vert\,
  \bar\theta_{12}^\textsc{reg}, \bar\theta_{13}, \Delta\bar{m}^2_{21} \big)
\end{multline}
with $\text{REG} = \text{LIGHT}$ or DARK.

\section{Results}
\label{sec:results}

In order to quantify the effect of the matter NSI on the present
oscillation parameter determination we have performed a set of 24
different analysis in the eleven-dimensional parameter space. Each
analysis corresponds to a different combination of observables. The
results of the long-baseline experiment MINOS are always included in
all the cases, so for convenience in what follows we define
$\chi^2_\text{OTHM} \equiv \chi^2_\text{OTH} + \chi^2_\text{MINOS}$.
To this we add $\chi^2_\text{LBL}$ with $\text{LBL} = \text{T2K}$,
NOvA, and T2K+NOvA, and for each of these three combinations we
present the results with and without the bounds of COHERENT. In
addition, we perform the analysis in four distinctive parts of the
parameter space: the solar octant ``REG'' being LIGHT or DARK, and the
mass ordering being normal (NO) or inverted (IO).

\begin{figure}\centering
  \includegraphics[width=0.99\textwidth]{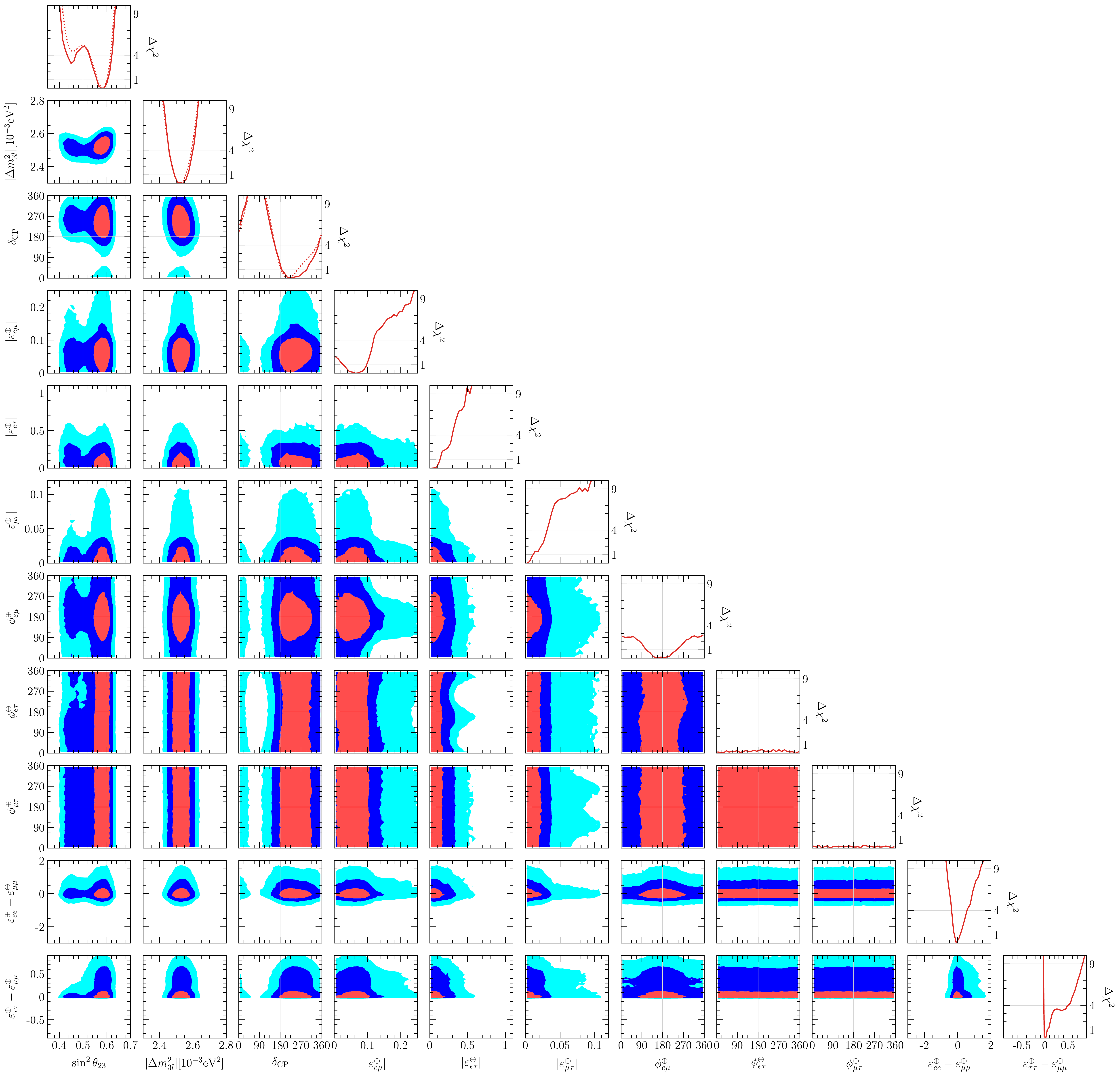}
  \caption{Global analysis of solar, atmospheric, reactor and
    accelerator oscillation experiments including the additional
    bounds from COHERENT experiment, in the LIGHT side of the
    parameter space and for Normal Ordering of the neutrino states.
    The panels show the two-dimensional projections of the allowed
    parameter space after marginalization with respect to the
    undisplayed parameters.  The different contours correspond to the
    allowed regions at $1\sigma$, $2\sigma$ and $3\sigma$ for
    2~degrees of freedom.  Note that as atmospheric mass-squared
    splitting we use $\Dmq_{3\ell} = \Dmq_{31}$ for NO. Also shown are
    the one-dimensional projections as a function of each
    parameter. For comparison we show as dotted lines the
    corresponding one-dimensional dependence for the same analysis
    assuming only standard $3\nu$ oscillation (\textit{i.e.}, setting
    all the NSI parameters to zero).}
  \label{fig:triangle-light}
\end{figure}

\begin{figure}\centering
  \includegraphics[width=0.99\textwidth]{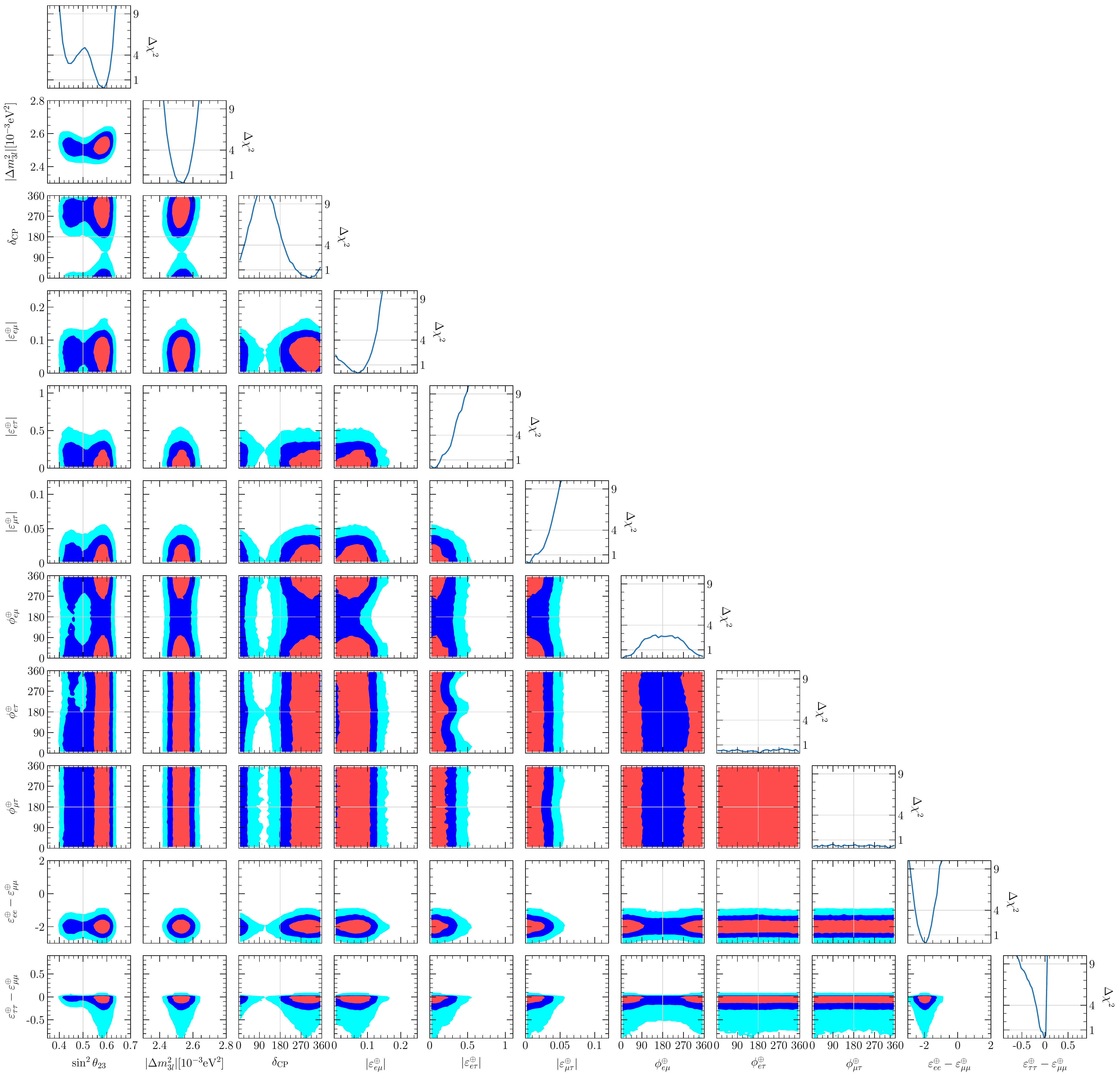}
  \caption{Same as Fig.~\ref{fig:triangle-dark} but for DARK-IO
    solution.  In this case $\Dmq_{3\ell} = \Dmq_{32}<0$ and we plot
    its absolute value. The regions and one-dimensional projections
    are defined with respect to the \emph{local} minimum in this
    sector of the parameter space.}
  \label{fig:triangle-dark}
\end{figure}

For illustration we show in Figs.~\ref{fig:triangle-light} and
Fig.~\ref{fig:triangle-dark} all the possible one-dimensional and
two-dimensional projections of the eleven-dimensional parameter space
accounting for the new CP violating phases, parametrized as
$\phi_{\alpha\beta} \equiv \arg(\Eps_{\alpha\beta}) =
\arg(\Eps_{\alpha\beta}^\oplus)$. In both figures we show the regions
for the GLOBAL analysis including both T2K and NOvA results and also
accounting for the COHERENT bounds. In Fig.~\ref{fig:triangle-light}
we present the results for the LIGHT sector and Normal Ordering, while
in Fig.~\ref{fig:triangle-dark} we give the regions corresponding to
the DARK sector and Inverted Ordering; in both cases the allowed
regions are defined with respect to the local minimum of each
solution. From these figures we can see that, with the exception of
the required large value of $\Eps_{ee}^\oplus - \Eps_{\mu\mu}^\oplus$
in the DARK solution, there is no statistically significant feature
for the $\Eps_{\alpha\beta}^\oplus$ parameters other than their
bounded absolute values, nor there is any meaningful information on
the $\phi_{\alpha\beta}$ phases.  The most prominent non-trivial
feature is the preference for a non-zero value of $\Eps_{e\mu}^\oplus$
at a $\Delta\chi^2 \sim 2$ level, associated with a $\phi_{e\mu}$
phase centered at the CP-conserving values $\pi$ ($0$) for the LIGHT
(DARK) solution.  More on this below.

\begin{figure}\centering
  \includegraphics[width=\textwidth]{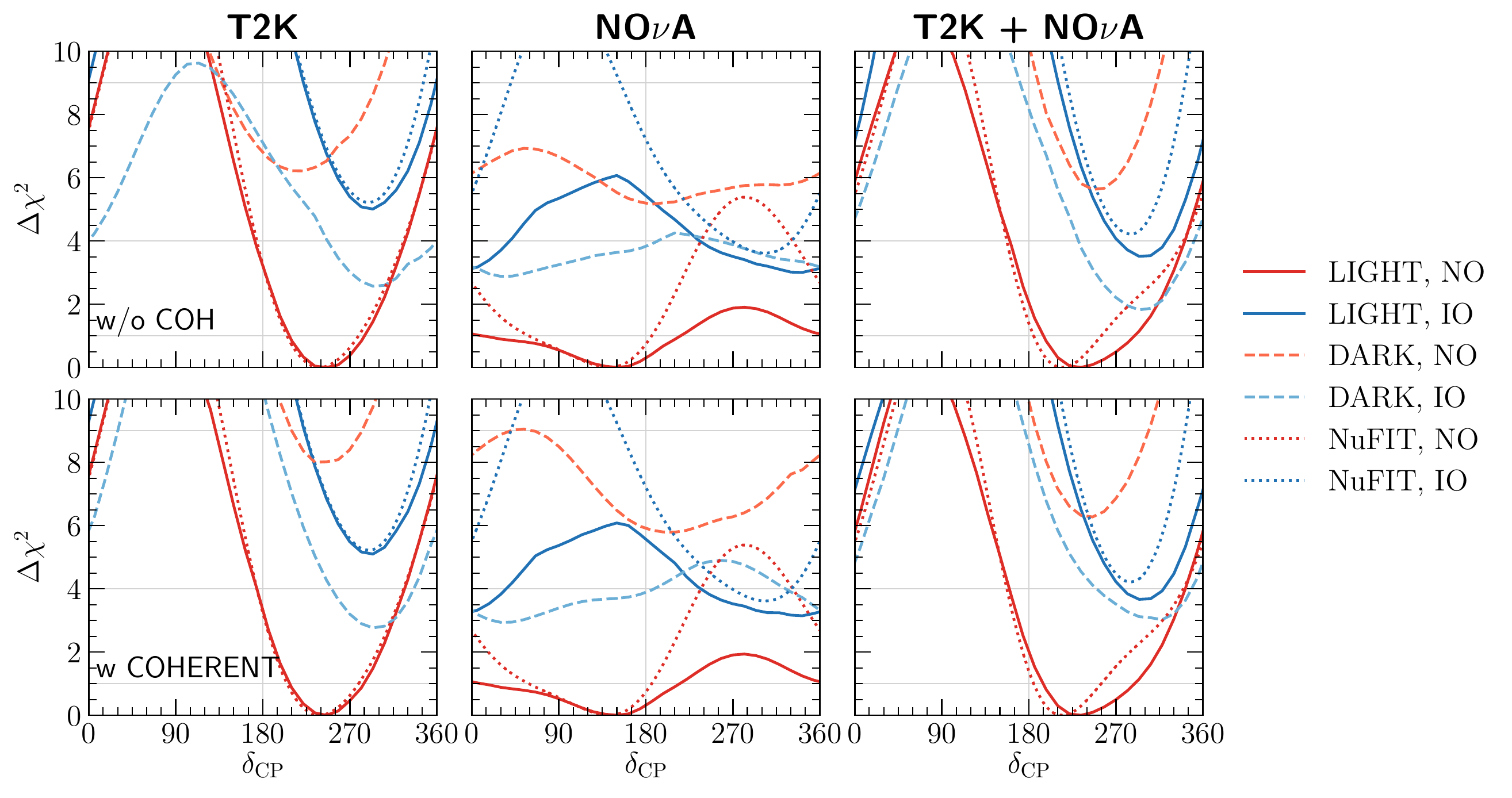}
  \caption{$\Delta\chi^2_\text{GLOB}$ as a function of
    $\delta_\text{CP}$ after marginalizing over all the undisplayed
    parameters, for different combination of experiments. In the upper
    panels we include $\text{SOLAR} + \text{KamLAND} + \text{MBL-REA}
    + \text{MINOS}$ to which we add T2K (left), NOvA (center) and
    $\text{T2K} + \text{NOvA}$ (right). The corresponding lower panels
    also include the constraints from COHERENT. The different curves
    are obtained by marginalizing within different regions of the
    parameter space, as detailed in the legend. See text for details.}
  \label{fig:chi2dcpnsi}
\end{figure}

In order to quantify the effect of the matter NSI on the present
determination of $\delta_\text{CP}$ and the mass ordering we plot in
Fig.~\ref{fig:chi2dcpnsi} the one-dimensional
$\chi^2(\delta_\text{CP})$ function obtained from the above
$\chi^2_\text{GLOB,REG}$ after marginalizing over the ten undisplayed
parameters. In the left, central and right panels we focus on the
GLOBAL analysis including T2K, NOvA, and T2K+NOvA respectively.  The
upper (lower) panels corresponds to the results without (with) the
inclusion of the bounds from COHERENT. In each panel we plot the
curves obtained marginalizing separately in NO (red curves) and IO
(blue curves) and within the $\text{REG} = \text{LIGHT}$ (full lines)
and $\text{REG} = \text{DARK}$ (dashed) regions. For the sake of
comparison we also plot the corresponding $\chi^2(\delta_\text{CP})$
from the $3\nu$ oscillation analysis with the SM matter potential
(labeled ``NuFIT'' in the figure).

For what concerns the analysis which includes T2K but not NOvA,
\textit{i.e.}, the left panels in Fig.~\ref{fig:chi2dcpnsi}, we find
that:
\begin{itemize}
\item The statistical significance of the hint for a non-zero
  $\delta_\text{CP}$ in T2K is robust under the inclusion of the
  NSI-induced matter potential for the most favored solution
  (\textit{i.e.}, LIGHT-NO), as well as for LIGHT-IO.  This statement
  holds irrespective of the inclusion of the bounds from COHERENT.

\item The $\Delta\chi^2$ for DARK solutions exhibits the expected
  inversion of the ordering as well as the $\delta_\text{CP} \to \pi -
  \delta_\text{CP}$ transformation when compared with the LIGHT
  ones. This is a consequence of the NSI-mass-ordering degeneracy
  discussed in Eqs.~\eqref{eq:osc-deg} and~\eqref{eq:NSI-deg}.

\item We notice that $\Delta\chi^2_\text{min,OTHM+T2K,DARK,IO} \neq
  \Delta\chi^2_\text{min,OTHM+T2K,LIGHT,NO}$ because of the breaking
  of the NSI-mass-ordering degeneracy in the analysis of solar
  experiments as a consequence of the sizeable variation of the
  chemical composition of the matter crossed by solar neutrinos along
  their path. However as we see in the upper left panel
  \begin{equation}
    \chi^2_\text{min,OTHM+T2K,DARK,IO}
    - \chi^2_\text{min,OTHM+T2K,LIGHT,NO} \simeq 2.5 < 3.15
  \end{equation}
  so the DARK solutions become less disfavored when T2K is included
  (see Eq.~\eqref{eq:dclmad}).  This suggests that the DARK-IO
  solution can provide a perfect fit to T2K data, \textit{i.e.}, there
  is an almost total loss of sensitivity to the ordering in T2K.
  Indeed what the inequality above shows is that within the allowed
  DARK parameter space it is possible to find areas where the fit to
  T2K-only data for IO are slightly better than the fit for NO in the
  LIGHT sector (and than NO oscillations without NSI).  For the same
  reason we also find that
  \begin{equation}
    \begin{aligned}
      \chi^2_\text{min,OTHM+T2K,DARK,NO}
      - \chi^2_\text{min,OTHM+T2K,DARK,IO} &\simeq 3.6   
      \\
      \chi^2_\text{min,OTHM+T2K,LIGHT,IO}
      - \chi^2_\text{min,OTHM+T2K,LIGHT,NO} &\simeq 5.1    
    \end{aligned}
      \Bigg\rbrace \Rightarrow 3.6 < 5.1\,,
     \end{equation}
  because there are DARK solutions with NO which give a ``less bad''
  fit than the degenerate of the LIGHT-IO minimum.  For example, in
  DARK-NO we find that the best fit value for $\Dmq_{31}$ can be
  slightly larger than the best-fit $|\Dmq_{32}|$ in LIGHT-IO, which
  leads to a slightly better agreement with the results on $\Dmq_{31}$
  from MBL reactors.  These solutions, however, involve large NSI
  parameters, in particular $\Eps_{e\mu}$ and $\Eps_{e\tau}$, which
  are disfavored by COHERENT.  Consequently in the lower left panel we
  read an slightly higher $\Delta\chi^2_\text{min,OTHM+T2K,DARK,IO} \simeq 3.$
  and a very similar difference in the quality of the fits between the
  orderings
  in LIGHT and DARK regions (with different sign of course), so
  $\chi^2_\text{min,OTHM+T2K,DARK,NO} -
  \chi^2_\text{min,OTHM+T2K,DARK,IO} \simeq 5.1$.

\item For the same reason, without including COHERENT, the statistical
  significance of the hint of CP violation in T2K is reduced for the
  DARK solutions with respect to the LIGHT ones. We find that CP
  conservation (CPC), that is, a fit with all phases either zero or
s  $\pi$, lies at
  \begin{equation}
    \begin{aligned}
      \chi^2_\text{CPC,OTHM+T2K,DARK,IO}
      - \chi^2_\text{min,OTHM+T2K,DARK,IO} &\simeq 1.5 
      \\
      \chi^2_\text{CPC,OTHM+T2K,LIGHT,NO}
      - \chi^2_\text{min,OTHM+T2K,LIGHT,NO} &\simeq 3.5 
    \end{aligned}
    \Bigg\rbrace \Rightarrow 1.5 < 3.5 \,.
  \end{equation}
  
\item However we still find that even without COHERENT
  \begin{equation}
    \begin{aligned}
      \chi^2_\text{CPC,OTHM+T2K,LIGHT,NO} &\simeq
      \chi^2_\text{OTHM+T2K,LIGHT,NO}(\delta_\text{CP}=\pi) \,,
      \\
      \chi^2_\text{CPC,OTHM+T2K,DARK,IO} &\simeq
      \chi^2_\text{OTHM+T2K,DARK,IO}(\delta_\text{CP}=0) \,.
    \end{aligned}
  \end{equation}
  So the CL for CPC as naively read from the curves of
  $\delta_\text{CP}$ still holds, or what is the same, there is no
  leptonic CP violation ``hidden'' when there is no CP violation from
  $\delta_\text{CP}$.
\end{itemize}

For the global combination including NOvA without T2K (central panels
in Fig.~\ref{fig:chi2dcpnsi}) we notice that:
\begin{itemize}
\item The sensitivity to $\delta_\text{CP}$ diminishes with respect to
  that of the oscillation only analysis both in the LIGHT and DARK
  sectors.  The constraints from COHERENT have a marginal impact on
  this.

\item Within the DARK sector, IO is the best solution as expected from
  the NSI/mass-ordering degeneracy, but it is still disfavored at
  $\Delta\chi^2_\text{min,OTHM+NOvA,DARK,IO}\sim 3$ because of
  SOLAR+KamLAND, Eq.~\eqref{eq:dclmad}. By chance, this happens to be
  of the same order of the difavoring of IO in the pure oscillation
  analysis, $\Delta\chi^2_\text{min,OTHM+NOvA,OSC}\sim 3.5$ (although
  the physical effect responsible for this is totally different).
\end{itemize}

In the global analysis including both T2K and NOvA (right panels in
Fig.~\ref{fig:chi2dcpnsi}) we find qualitatively similar conclusions
than for the analysis without NOvA, albeit with a slight washout of
the statistical significance for both $\delta_\text{CP}$ and the
disfavoring of IO due to the tensions between T2K and NOvA. Such
washout is already present in the oscillation-only analysis and within
the LIGHT sector it is only mildly affected by the inclusion of
NSI. However one also observes that in the favored solution, LIGHT-NO,
maximal $\delta_\text{CP} = 3\pi/2$ is more allowed than without
NSI. This happens because, as mentioned above, in NOvA the presence of
NSI induces a loss of sensitivity on $\delta_\text{CP}$, so in the
global analysis with both T2K and NOvA the behavior observed in T2K
dominates.

In the global analysis there remains, still, the DARK-IO solution at
\begin{equation}
  \chi^2_\text{min,GLOB,DARK,IO}
  - \chi^2_\text{min,GLOB,LIGHT,NO} = 2~(3)
\end{equation}
without (with) including COHERENT.

\begin{figure}\centering
  \includegraphics[width=\textwidth]{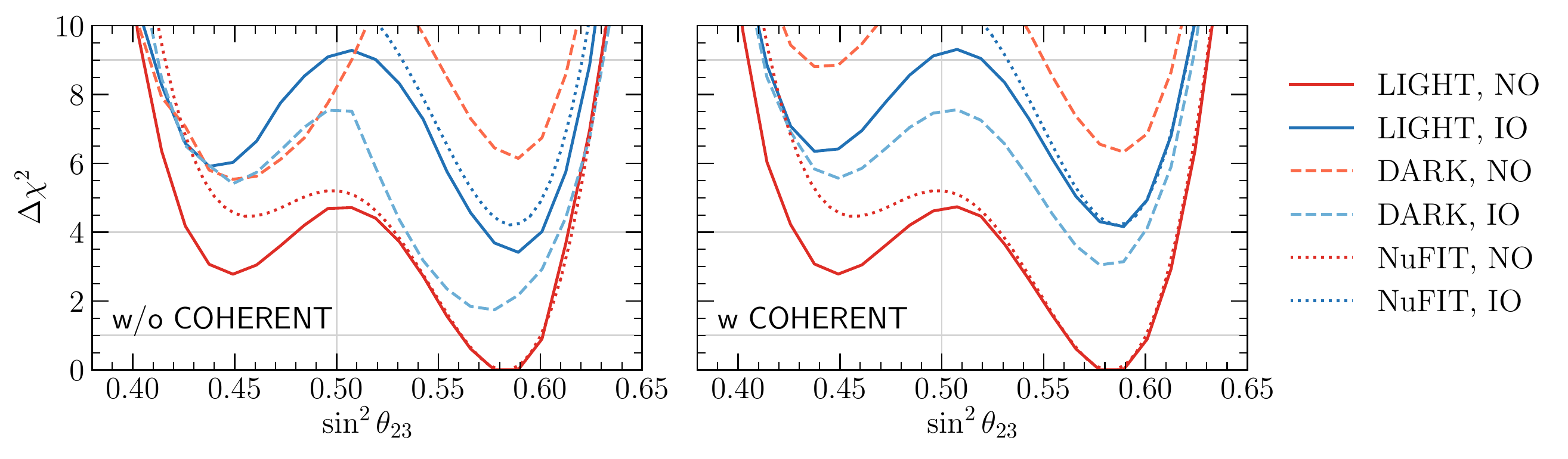}
  \caption{$\Delta\chi^2_\text{GLOB}$ as a function of
    $\sin^2\theta_{23}$ after marginalizing over all other parameters
    for the GLOBAL combination of oscillation experiments without
    (with) COHERENT in the left (right) panels.  The different curves
    correspond to marginalization within the different regions of the
    parameter space, as detailed in the legend. See text for details.}
  \label{fig:t23}
\end{figure}

The status of the non-maximality and octant determination for
$\theta_{23}$ is displayed in Fig.~\ref{fig:t23} where we show the
one-dimensional $\chi^2(\sin^2\theta_{23})$ obtained from
$\chi^2_\text{GLOB,REG}$ including both T2K and NOvA and after
marginalizing over all the undisplayed parameters (so these are the
corresponding projections to the left panels in
Fig.~\ref{fig:chi2dcpnsi}).  In the left (right) panel of
Fig.~\ref{fig:t23} we show the results without (with) COHERENT.
As
seen in the figure the global analysis including NSI for both
orderings, for both LIGHT and DARK sectors, and irrespective of the
inclusion of COHERENT, still disfavors the maximal $\theta_{23}=\pi/4$
at a CL $\sim 2\sigma \div 2.5\sigma$.  The main effect of the
generalized NSI matter potential on $\theta_{23}$ is the relative
improvement of the CL for the first octant, which, compared to the
best fit in the second octant is now at less than $2\sigma$. For
example for the two most favored solutions it is now at
\begin{equation}
  \begin{aligned}
    \chi^2_\text{GLOB,LIGHT,NO} (\theta_{23} <45^\circ)
    -\chi^2_\text{GLOB,LIGHT,NO} (\theta_{23} >  45^\circ)
    &\gtrsim 3~(3) \,,
    \\
    \chi^2_\text{GLOB,DARK,IO} (\theta_{23} < 45^\circ)-\chi^2_\text{GLOB,DARK,IO} (\theta_{23} > 45^\circ)
    &\gtrsim 3.8~(2.8)
  \end{aligned}
\end{equation}
for the analysis without (with) COHERENT.  This is so because the
disfavoring of the first octant in the global oscillation-only
analysis is driven by the excess of appearance events in NOvA.  These
events can now be fitted better with $\theta_{23}$ in the first octant
when including a non-zero $\Eps_{e\mu}$ to enhance the $\nu_\mu \to
\nu_e$ flavor transition probability.  For this reason, as seen in the
left panel in Fig.~\ref{fig:t23} in the case of DARK-NO without
COHERENT, the first octant becomes favored.  But, as mentioned above,
these solutions required large NSI which are disfavored by COHERENT
and therefore once this is included in the analysis, the second octant
becomes the best fit in all regions of parameter space.

\section{Summary}
\label{sec:summary}

In this work we have extended the analysis in
Ref.~\cite{Esteban:2018ppq} to account for the effect of NSI affecting
neutrino propagation in matter on the observables sensitive to
leptonic CP violation and to the mass ordering. We have quantified the
robustness of the present hints for these effects in the presence of
NSI as large as allowed by the global oscillation analysis itself. We
conclude that the CL for the preference for a CKM-like CP phase close
to $3\pi/2$ in T2K, which is the one that drives the preference in the
global analysis, remains valid even when including all other phases in
the extended scenario. On the contrary the preference for NO in LBL
experiments is totally lost when including NSI as large as allowed
by the global analysis because of the intrinsic NSI/mass-ordering
degeneracy in the Hamiltonian which implies the existence of an
equally good fit to LBL results with IO and reversed octant of
$\theta_{12}$ and $\delta_\text{CP}\rightarrow \pi - \delta_\text{CP}$
(so in this solution the favored $\delta_{CP}$ is also close to
$3\pi/2$).  In the global analysis the only relevant breaking of this
degeneracy comes from the composition dependence of the matter
potential in the Sun which disfavors the associated LMA-D with CL
below $2\sigma$. Finally, we have also  studied the effect of NSI in
the status of the non-maximality and octant determination for
$\theta_{23}$ and find that for both
orderings, for both LIGHT and DARK sectors, and irrespective of the
inclusion of COHERENT,  maximal $\theta_{23}=\pi/4$ is still disfavoured 
in the global fit at a CL $\sim 2\sigma \div 2.5\sigma$. Including
NSI, however, results into a decrease of the preference for the
second octant  with respect to the first octant to less than 2 $\sigma$.

\section*{Acknowledgements}

This work is supported by USA-NSF grant PHY-1620628, by EU Networks
FP10 ITN ELUSIVES (H2020-MSCA-ITN-2015-674896) and INVISIBLES-PLUS
(H2020-MSCA-RISE-2015-690575), by MINECO grant FPA2016-76005-C2-1-P
and MINECO/FEDER-UE grants FPA2015-65929-P and FPA2016-78645-P, by
Maria de Maetzu program grant MDM-2014-0367 of ICCUB, and by the
``Severo Ochoa'' program grant SEV-2016-0597 of IFT.
I.E.\ acknowledges support from the FPU program fellowship FPU15/03697.

\appendix

\section{Invariants for leptonic CP violation with NSI}
\label{sec:appendix}

In this appendix we derive a set of basis and rephasing invariants
that characterize leptonic CP violation in the presence of
non-standard neutrino interactions. We follow the methodology
introduced in Refs.~\cite{Bernabeu:1986fc, Gronau:1986xb} for
generalizing the construction of such invariants in quark sectors
first introduced for three generations in~\cite{Jarlskog:1985cw,
  Jarlskog:1985ht}.  Notice that we are working with Dirac neutrinos
which is all it is needed when interested in CP violation in neutrino
oscillations.

The relevant parts of the Lagrangian are:
\begin{equation}
  \begin{split}
    \mathcal{L} = & +
    i \bar{\nu}_{\alpha, L} \slashed{\partial} \nu_{\alpha, L} +
    i \bar{\nu}_{\alpha, R} \slashed{\partial} \nu_{\alpha, R} +
    i \bar{\ell}_{\alpha, L} \slashed{\partial} \ell_{\alpha, L} +
    i \bar{\ell}_{\alpha, R} \slashed{\partial} \ell_{\alpha, R}
    \\
    & -\frac{g}{\cos \theta_W} Z_\mu \bigg[ \frac{1}{2} \bar{\nu}_{\alpha,L}
      \gamma^\mu \nu_{\alpha, L} - \left(\frac{1}{2} + \sin^2
      \theta_W\right) \bar{\ell}_{\alpha, L} \gamma^\mu \ell_{\alpha, L} -
      \sin^2 \theta_W \bar{\ell}_{\alpha, L} \gamma^\mu \ell_{\alpha, R}
      \bigg]
    \\
    & +\bigg[- \frac{g}{\sqrt{2}} W_\mu^- \bar{\ell}_{\alpha, L}
      \gamma^\mu \nu_{\alpha, L}
      - \sum_{\alpha \beta}  M^\nu_{\alpha \beta} \bar{\nu}_{\alpha, L}
      \nu_{\beta, R} - \sum_{\alpha \beta}
      M^\ell_{\alpha \beta} \bar\ell_{\alpha,L} \ell_{\beta, R}
      + \text{h.c.} \bigg]
    \\
    & -2 \sqrt{2} G_F \bigg[ \sum_{\alpha \beta} \Eps_{\alpha \beta}
      \left(\bar{\nu}_\alpha \gamma^\mu P_L \nu_\beta \right) \bigg]
    \bigg[\sum_{f, P} \xi^{f,P} \left(\bar{f}\gamma_\mu P f\right) \bigg]
  \end{split}
\end{equation}
with $\alpha \in \{e,\mu,\tau\}$ and $i \in \{1,2,3\}$.

Unphysical flavor basis rotations are given by the following field
transformations
\begin{align}
  \ell_L & \stackrel{\text{flavor}}{\longrightarrow} P^L \, \ell_L \,,
  &&
  \nu_L \stackrel{\text{flavor}}{\longrightarrow} P^L \, \nu_L \,,
  \\
  \ell_R & \stackrel{\text{flavor}}{\longrightarrow} P^{\ell, R} \, \ell_R
  \,,
  &&
  \nu_R \stackrel{\text{flavor}}{\longrightarrow} P^{\nu, R} \, \nu_R \,,
\end{align}
where all the $P$ unitary matrices. Correspondingly the matrices
with flavor indexes transform as
\begin{equation}
  M^\nu \stackrel{\text{flavor}}{\longrightarrow}
  P^{L \dagger} M^\nu P^{\nu, R} \,,
  \qquad
  M^\ell \stackrel{\text{flavor}}{\longrightarrow}
  P^{L \dagger} M^\ell P^{\ell, R} \,,
  \qquad
  \Eps \stackrel{\text{flavor}}{\longrightarrow}
  P^{L \dagger} \Eps P^{L} \,.
\end{equation}
CP conjugation transforms the matrices with flavor indexses into
their complex conjugates,
\begin{equation}
  M^\nu \stackrel{\text{CP}}{\rightarrow} M^{\nu *} \,,
  \qquad
  M^\ell \stackrel{\text{CP}}{\rightarrow} M^{\ell *} \,,
  \qquad
  \Eps \stackrel{\text{CP}}{\rightarrow} \Eps^* \,.
\end{equation}
So clearly the CP transformation will be unphysical if (and only if)
it is equivalent to some flavor rotation. That is, there is CP
conservation if and only if there exists a set of unitary matrices
$\{P^L, P^{\nu, R}, P^{l, R}\}$ such that
\begin{equation}
  P^{L \dagger} \Eps P^{L}  = \Eps^* \,,
  \qquad
  P^{L \dagger} M^\nu P^{\nu, R}  = M^{\nu *} \,,
  \qquad
  P^{L \dagger} M^\ell P^{\ell, R}  = M^{\ell *} \,.
\end{equation}
Since given a matrix $A$, $A A^\dagger$ determines $A$ up to unitary
rotations one can work with the ``squares'' of the mass matrices
instead and find that there is CP conservation if and only if there
exists a unitary matrix $P$ such that
\begin{equation}
  \label{eq:CPconsCondition}
  P^\dagger \Eps P  = \Eps^* \,,
  \qquad
  P^\dagger S^\nu P  = (S^\nu)^* \,,
  \qquad
  P^\dagger S^\ell P  = (S^\ell)^* \,, 
\end{equation}
with
$S^\ell =M^\ell M^{\ell \dagger}$ and $S^\nu = M^\nu M^{\nu \dagger}$
which in the charged fermion mass basis are
\begin{equation}
  S^\ell =
  \begin{pmatrix}
    m_e^2 & 0 & 0 \\
    0 & m_\mu^2 & 0 \\
    0 & 0 & m_\tau^2
  \end{pmatrix} \,,
  \qquad
  S^\nu  = U
  \begin{pmatrix}
    m_1 & 0 & 0 \\
    0 & m_2 & 0 \\
    0 & 0 & m_3
  \end{pmatrix}
  U^\dagger \,,
\end{equation}
where $U = U_\text{vac}$ is the leptonic mixing matrix. In this basis
the conditions \eqref{eq:CPconsCondition} read
\begin{align}
  \label{eq:CPconsElementsCondition1}
  \Im \left(S^\nu_{e\mu}\, S^\nu_{\mu\tau}\, S^\nu_{\tau e} \right) &= 0 \,,
  \\
  \Im \left( \Eps_{e\mu}\, \Eps_{\mu\tau}\, \Eps_{\tau e} \right) &= 0 \,,
  \\
  \label{eq:CPconsElementsCondition3}
  \Im \left( \Eps_{\alpha\beta} \,S^\nu_{\beta\alpha} \,\right) &= 0 \,.
\end{align}
Note that the last condition has to be fulfilled for $\{\alpha\beta\}
= \{e\mu\}, \{e,\tau\}, \{\mu,\tau\}$. However, since
\begin{equation}
    \label{eq:nonIndependentCondition}
  \Eps_{\mu\tau}\, S^\nu_{\tau\mu} =
  \frac{\big(\Eps_{e\mu}\, \Eps_{\mu\tau}\, \Eps_{\tau e}\big)
    \big(S^\nu_{e\mu}\, S^\nu_{\mu\tau}\, S^\nu_{\tau e}\big)^*
    \big(\Eps_{e\tau} \,S^\nu_{\tau e}\big)
    \big(\Eps_{e\mu}\, S^\nu_{\mu e}\big)^*}
    {\big|\Eps_{e\mu}\big|^2 \big|\Eps_{e\tau}\big|^2
      \big|S^\nu_{e\mu}\big|^2 \big|S^\nu_{e\tau}\big|^2}
\end{equation}
there are only four independent conditions.

Using the projector technique~\cite{Jarlskog:1987zd} the four
conditions can be expressed in a basis-invariant form. For example as:
\begin{align}
  \Im\Tr\left[ (S^\ell)^2\, (S^\nu)^2\, S^\ell\, S^\nu \right]
  &= \frac{2}{i} \Det [S^\ell, S^\nu] = 0
  \\
  \Im\Tr\left[ (S^\ell)^2\, (\Eps)^2\, S^\ell\, \Eps \right]
  &= \frac{2}{i} \Det [S^\ell, \Eps]=0\,,
  \\
  \Im\Tr\left[ S^\nu\, S^\ell\,\Eps \right] &= 0\,,
  \\
  \Im\Tr\left[ S^\ell\, S^\nu\, (S^\ell)^2\, \Eps \right] &= 0 \,.
\end{align}
In the basis where the lepton mass matrix is diagonal these invariants read
\begin{align}
  \label{eq:appinv1}
  \Im\Tr\left[ (S^\ell)^2\, (S^\nu)^2\, S^\ell\, S^\nu \right]
  & = v(m_e, m_\mu,m_\tau)
  \Im\left[ S^\nu_{e\mu}\, S^\nu_{\mu\tau}\, S^\nu_{\tau e} \right]
  \\
  \label{eq:appinv2}
  \Im\Tr\left[ (S^\ell)^2\, (\Eps)^2\, S^\ell\, \Eps \right]
  & = v(m_e, m_\mu,m_\tau)
  \Im\left[ \Eps_{e\mu}\, \Eps_{\mu\tau}\, \Eps_{\tau e} \right] \,,
  \\
  \label{eq:appinv3}
  \begin{split}
    \Im \Tr\left[ S^\nu\, S^\ell\, \Eps \right]
    &= (m_\mu^2-m_e^2) \Im\left( S^\nu_{e\mu} \Eps_{\mu e} \right)
    + (m_\tau^2-m_e^2) \Im\left( S^\nu_{e\tau} \Eps_{\tau e} \right)
    \\
    & \quad + (m_\tau^2-m_\mu^2)
    \Im \left( S^\nu_{\mu\tau}\Eps_{\tau\mu} \right)
  \end{split}
  \\
  \label{eq:appinv4}
  \begin{split}
    \Im \Tr\left[ (S^\ell)^2\, (\Eps)^2\, S^\ell\,
      \Eps \right]
    &= m_e m_\mu (m_\mu^2-m_e^2)
    \Im\left( S^\nu_{e\mu} \Eps_{\mu e} \right)
    + m_e m_\tau (m_\tau^2-m_e^2)
    \Im\left( S^\nu_{e\tau} \Eps_{\tau e} \right)
    \\
    & \quad + m_\mu m_\tau(m_\tau^2-m_\mu^2)
    \Im \left( S^\nu_{\mu\tau} \Eps_{\tau\mu} \right)
  \end{split}
\end{align}
with $v(m_e,m_\mu,m_\tau) = (m_\tau^2-m_\mu^2) (m_\tau^2-m_e^2)
(m_\mu^2-m_e^2)$.

Written in this form, the conditions for which the four independent
phases are physically realizable becomes explicit, in particular the
requirement of the non-zero difference between all or some of the
charged lepton masses. Notice, however, that the amplitudes for the
neutrino flavor transition observables do not have to explicitly
display such dependence on the charged lepton mass, because they
correspond to transitions between an inital and a final state, each
associated with some specific charged lepton~\cite{Jarlskog:1986mm}.
Hence these basis invariants for leptonic CP-violating observables are
of limited applicability for expressing the neutrino oscillation
probabilities.

\bibliographystyle{JHEP}
\bibliography{references}

\end{document}